\pdfoutput=1

\documentclass[11pt,twoside,a4paper,cmspaper,final,collab]{cms-tdr}
\def\svnVersion{456817P}\def\cmsCernNoTag{CERN-EP-2018-005}\def\cmsCernDate{\today}\def\cmsMessage{Published in the Journal of High Energy Physics as \href{http://dx.doi.org/10.1007/JHEP03(2018)181}{\doi{10.1007/JHEP03(2018)181.}}}

\begin{document}\cmsNoteHeader{HIN-16-005}

\hyphenation{had-ron-i-za-tion}
\hyphenation{cal-or-i-me-ter}
\hyphenation{de-vices}
\RCS$HeadURL: svn+ssh://svn.cern.ch/reps/tdr2/papers/HIN-16-005/trunk/HIN-16-005.tex $
\RCS$Id: HIN-16-005.tex 451701 2018-03-19 15:50:43Z mnguyen $

\providecommand{\NA}{\ensuremath{\text{---}}}
\providecommand{\HYDJET}{\textsc{hydjet}\xspace}
\newcommand{\PbPb}{\ensuremath{\mathrm{PbPb}}\xspace}
\newcommand{\pp}{\ensuremath{\Pp\Pp}\xspace}
\newcommand{\xj}{\ensuremath{x_\mathrm{J}}\xspace}
\newcommand{\meanxj}{\ensuremath{\langle x_\mathrm{J}\rangle}\xspace}
\newcommand{\dphi}{\ensuremath{\Delta\phi}\xspace}
\newcommand{\JEWEL} {{\textsc{jewel}}\xspace}

\cmsNoteHeader{HIN-16-005}
\title{Comparing transverse momentum balance of \cPqb{} jet pairs in \pp and \PbPb collisions at $\sqrtsNN = 5.02$\TeV}

\date{\today}

\abstract{
The transverse momentum balance of pairs of back-to-back \cPqb{} quark jets in \PbPb and \pp collisions recorded with the CMS detector at the LHC is reported.  The center-of-mass energy in both collision systems is 5.02\TeV per nucleon pair.  Compared to the \pp collision baseline, \cPqb{} quark jets have a larger imbalance in the most central \PbPb collisions, as expected from the jet quenching effect.  The data are also compared to the corresponding measurement with inclusive dijets.  In the most central collisions, the imbalance of \cPqb{} quark dijets is comparable to that of inclusive dijets.
}

\hypersetup{%
pdfauthor={CMS Collaboration},%
pdftitle={Comparing transverse momentum balance of b jet pairs in pp and PbPb collisions at sqrt(s[NN]) = 5.02 TeV},%
pdfsubject={CMS},%
pdfkeywords={CMS, physics}}

\maketitle

\section{Introduction}
\label{sec:intro}

Jets are sensitive probes of final-state effects in heavy ion collisions.  The jet quenching phenomenon is understood to arise from the interaction of hard-scattered partons with the quark-gluon plasma produced in such collisions~\cite{Qin:2015srf}.   The first observable used to probe this phenomenon at the LHC was the transverse momentum (\pt) balance of back-to-back jets~\cite{Aad:2010bu, Chatrchyan:2011sx, Khachatryan:2015lha, Aaboud:2017eww}.  Quenching imparts a net imbalance to dijets that exceeds the imbalance from QCD radiation in vacuum, as measured in \pp collisions.  This additional imbalance is expected based on the difference of the in-medium path-length traversed by the two jets.  However, jet-by-jet fluctuation of the quenching may also play a role, and could even be dominant~\cite{Milhano:2015mng}.

The dependence of quenching on the type of parton that initiates the jet may provide insight into the underlying dynamics.  Such a dependence could arise directly from the interaction of the initiating parton with the medium.  For example, radiative loss via gluon bremsstrahlung is expected to be larger for jets initiated by gluons than for those from quarks.  Furthermore, for heavy quarks, radiation is expected to be suppressed in the direction of propagation~\cite{Dokshitzer:2001zm}.
A dependence could also arise less directly, via the medium interactions of subleading partons in the shower.  For models in which quenching depends on the shower multiplicity, \eg, \JEWEL~\cite{Zapp:2008gi, Milhano:2015mng}, the relatively larger average parton multiplicity of gluon-initiated jets would lead to a larger quenching effect.

In general, the type of parton that initiates the jet is difficult to determine experimentally.  A notable exception are jets produced by the fragmentation of bottom quarks.  The corresponding b hadron may be identified, for example, by the presence of a soft lepton or a displaced vertex inside the jet.  The latter strategy was pursued in the CMS measurement of the \cPqb{} quark jet (``\cPqb{} jet'') spectra and the corresponding nuclear modification factor in \PbPb collisions at a nucleon-nucleon center of mass energy of $\sqrtsNN = 2.76$\TeV~\cite{Chatrchyan:2013exa}.  However, there is a potential ambiguity in that measurement.  Bottom quarks may be produced not only directly in the hard scattering, but also in the subsequent splitting of gluons into \cPqb{} quark pairs.  Jets associated with b hadrons may contain a significant contribution from gluon splitting, both from gluons that participate directly in the hard scattering, as well as those that arise from final-state radiation in the parton shower process.

One way to suppress the contribution of gluon splitting, which tends to produce pairs of \cPqb{} quarks with a relatively small opening angle, is to look at pairs of \cPqb{} jets that are back-to-back in azimuth. As shown in the Appendix, this configuration enhances the contribution from primary \cPqb{} quarks,  typically produced via the reaction $\Pg\Pg \to \bbbar$.  The \pt balance of such \cPqb{} jets may then be compared with those of inclusive (i.e., nontagged) dijets.  This paper presents the first measurement of the \pt balance of \cPqb{} jet pairs (``\cPqb{} dijets'') in \PbPb, using collisions recorded at $\sqrtsNN = 5.02$\TeV. The \cPqb{} dijet data are compared with that of inclusive dijets to search for a possible dependence of the \pt balance on the species of the initiating partons.

\section{The CMS detector}
\label{sec:detector}

The central feature of the CMS apparatus is a superconducting solenoid of 6\unit{m} internal diameter, providing a magnetic field of 3.8\unit{T}. Within the solenoid volume are a silicon pixel and strip tracker, a lead tungstate crystal electromagnetic calorimeter, and a brass and scintillator hadron calorimeter, each composed of a barrel and two endcap sections. Forward calorimeters extend the pseudorapidity coverage provided by the barrel and endcap detectors over the range of about $3 < \eta < 5$. Muons are detected in gas-ionization chambers embedded in the steel flux-return yoke outside the solenoid. A more detailed description of the CMS detector, together with a definition of the coordinate system used and the relevant kinematic variables, can be found in Ref.~\cite{Chatrchyan:2008zzk}.

Events of interest are selected using a two-tiered trigger system~\cite{Khachatryan:2016bia}. The first level, composed of custom hardware processors, uses information from the calorimeters and muon detectors. The second level, known as the high-level trigger, consists of a farm of processors running a version of the full event reconstruction software optimized for fast processing.

\section{Event and object selection}
\label{sec:selection}

 This analysis is performed using \PbPb and \pp data recorded in 2015 at a center-of-mass energy per nucleon pair of $\sqrtsNN = 5.02\TeV$.  The \PbPb and \pp samples correspond to integrated luminosities of 404\mubinv and 25.8\pbinv, respectively.
Events were selected using single-jet triggers in both \pp and \PbPb collisions.  The jet triggers used in this analysis are fully efficient with respect to the offline leading jet selection of $\pt > 100 \GeV$.
For \PbPb collisions, b tagging algorithms are applied at the high-level trigger to reduce the data volume.   This is achieved by performing a simplified version of the charged-particle tracking and vertex reconstruction in regions of the detector delineated by high-\pt jets.   The efficiency of the online b tagging with respect to the corresponding offline algorithm is evaluated using single-jet triggers, and lies in the range of 70--90\%, depending on collision centrality.

To reject noncollision processes such as beam-gas interactions, events are required to have at least one reconstructed primary vertex and to deposit an energy of at least 3\GeV in at least 3 towers in each of the two forward calorimeters.  The forward calorimeters are also used to estimate the collision centrality, evaluated as a percentile of the total inelastic hadronic cross section, with the most central event corresponding to a centrality of 0\%.

The anti-\kt algorithm~\cite{Cacciari:2008gp} is used to cluster jets from objects produced by the CMS particle-flow algorithm~\cite{Sirunyan:2017ulk}, which combines information from the various subdetector systems. A radius parameter of $R = 0.4$ is used.  In \PbPb collisions, the heavy ion background is subtracted event-by-event with an algorithm that is a variant of an iterative ``noise/pedestal subtraction'' technique~\cite{Kodolova:2007hd}.   The jet energy is calibrated as a function of the $\eta$ and \pt following the procedure described in Ref.~\cite{Chatrchyan:2011ds}.

The identification of \cPqb{} jets is achieved using the ``combined secondary vertex'' (CSV) discriminator.  This algorithm takes as input a number of properties of the reconstructed secondary vertex (SV), such as its displacement, the number of associated tracks, and their invariant mass (with the assumption that the tracks are originated by charged pions).  For events in which no SV is properly reconstructed, the displacement of selected tracks is used.  Details of the b tagging algorithms, and tracking and vertexing in general, can be found in Refs.~\cite{Chatrchyan:2012jua} and ~\cite{Chatrchyan:2014fea}, respectively. Simulated data samples produced with \GEANTfour~\cite{Agostinelli:2002hh} are used to evaluate the b tagging performance and derive various corrections. These samples are generated with \PYTHIA version 6.423~\cite{Sjostrand:2006za}, tune Z2~\cite{Field:2011iq}.  To compare with \PbPb data, \PYTHIA events are embedded in an underlying event produced with the \HYDJET generator, version 1.9~\cite{Lokhtin:2005px}.

\begin{figure}[!h]
  \centering
    \includegraphics[width=0.6\textwidth]{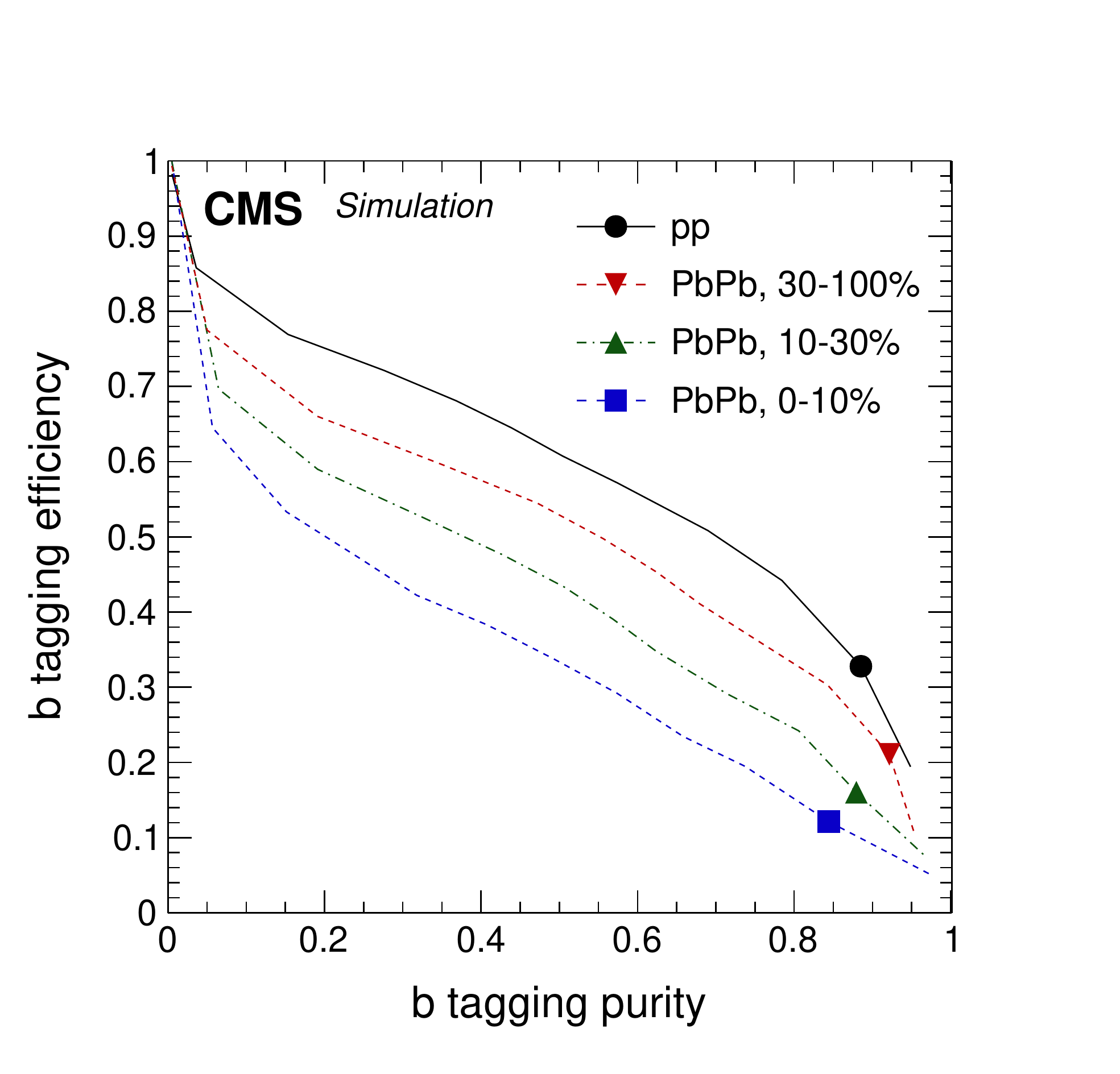}
 \caption[ROC curve]{The \cPqb{} dijet purity vs. efficiency as a function of the value of the selection on the CSV discriminator in simulation.  The same CSV selection is applied to both jets. Several different centrality intervals of \PbPb, as well as \pp collisions,  are shown, as indicated in the legend.  The closed symbols indicate the working point used in this analysis.  }
\label{fig:roc}
\end{figure}

  The performance of the CSV algorithm to identify \cPqb{} jet pairs offline is shown in Fig.~\ref{fig:roc}.  The efficiency and purity are evaluated in simulation as a function of the \cPqb-tagging selection variable for \pp and \PbPb collisions for different centrality intervals.  A tight selection on the CSV discriminator is applied in this analysis, as indicated in Fig~\ref{fig:roc}, leading to a purity in the range of 85--95\% for \cPqb{} dijets, with an efficiency in the range of about 10--35\%, depending on collision system and centrality.   The degradation of the performance with increasing centrality corresponds to a larger mistagging rate for fixed b tagging efficiency, as also observed in Ref.~\cite{Chatrchyan:2013exa}.  These jets are mistagged primarily due to vertices from false track combinations.

\section{Data analysis}
\label{sec:analysis}

The \pt balance of dijets is measured using the leading and subleading jets.  This balance is quantified by the ratio of the subleading to leading jet \pt, denoted \xj.  Dijets are selected from the two highest \pt jets within a window of  $\abs{\eta} < 1.5$.  The \pt of the leading and the subleading jets are required to be above 100 and 40\GeV, respectively.  This asymmetric \pt selection is chosen to ensure sensitivity to quenching effects.  The subleading jet threshold of 40\GeV is chosen to keep the subleading jet-finding efficiency reasonably high, as will be described below.  The leading jet threshold of 100\GeV is a compromise between statistical precision, on one hand, and maintaining a large lever-arm with the subleading jet, on the other.  For the case of \cPqb{} dijets, the leading and subleading jets are chosen prior to \cPqb-tagging selection.  By restricting the analysis to the two highest \pt jets in the event, the contribution from gluon splitting processes is significantly suppressed.

\begin{figure}[h!]
  \centering
    \includegraphics[width=1.0\textwidth]{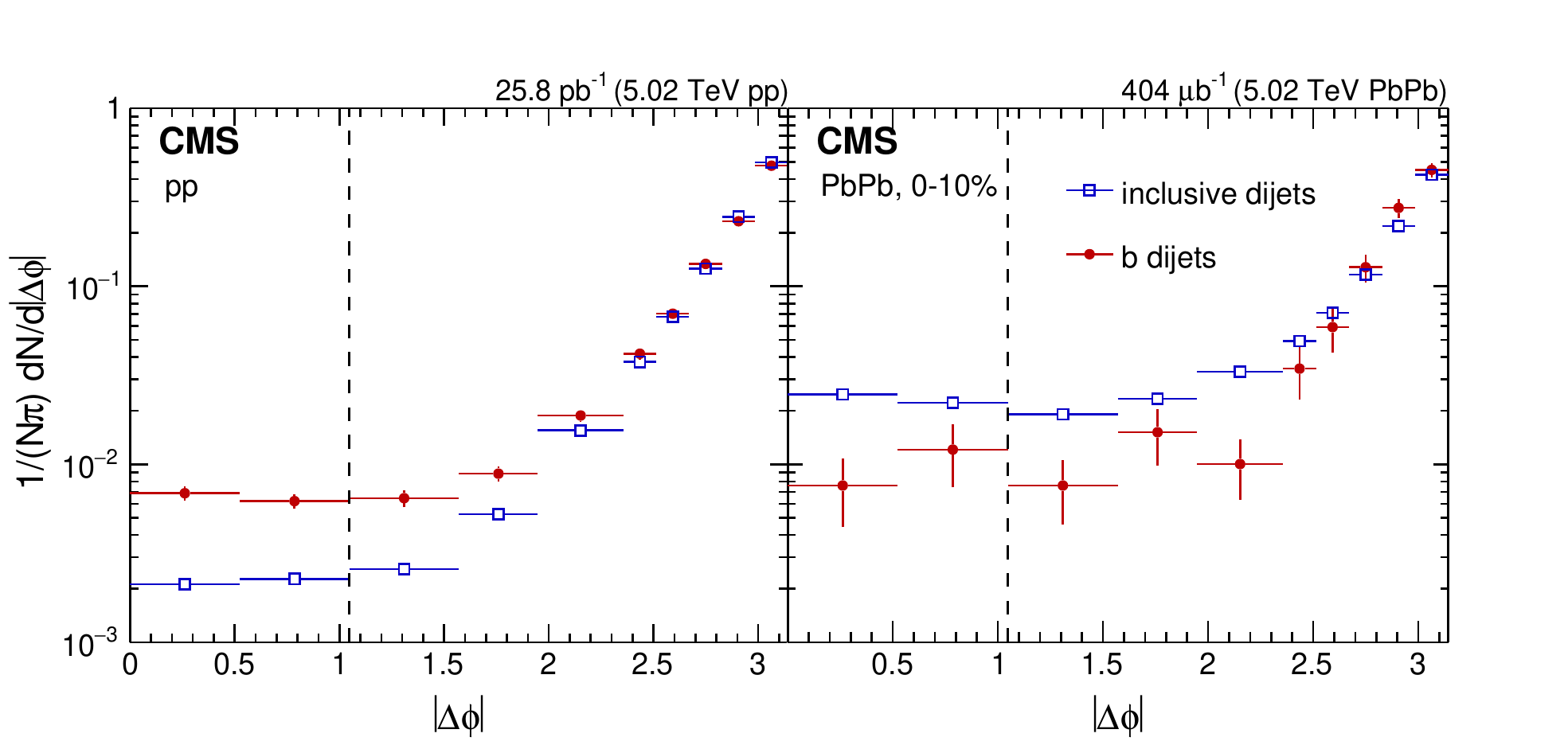}
 \caption[$\Delta \phi$ correlations]{ Distributions of the azimuthal opening angle ($\Delta \phi$) between the leading and subleading jets for \pp (left) and central (0--10\%) \PbPb collisions (right) for inclusive dijets and \cPqb{} dijets.  The small-angle region ($\abs{\Delta \phi}< \pi/3$), the boundary of which is indicated by a dashed line, is used to evaluate the combinatorial contribution in \PbPb collisions. The vertical bars represent statistical uncertainties, while the horizontal bars represent the bin widths.
   \label{fig:dphi}
 }
\end{figure}

Pairs of jets from a single hard scattering are referred to as ``signal'' pairs.  To enhance the contribution of such pairs, the jets are required to be back-to-back in azimuthal opening angle with the selection of $\abs{\Delta \phi} > 2\pi/3$.  The $\Delta \phi$ distributions in \pp collisions for inclusive dijets and dijets for which both the leading and subleading jets are b tagged are shown in the left panel of Fig.~\ref{fig:dphi}.  The b-tagged dijets show a more pronounced tail at small \dphi, which comes from a larger contribution of 3-jet topologies, as further discussed in the Appendix.  The \dphi  distributions in central (0--10\%) \PbPb collisions are shown in the right panel of Fig.~\ref{fig:dphi}.  For inclusive dijets, an increased contribution (compared to \pp collisions) at small \dphi  arises from pairs of jets that are not from the same nucleon-nucleon interaction.   These combinatorial jet pairs tend to bias the \xj distribution towards low values, i.e., towards large imbalance.
To subtract this contribution from the selected dijet pairs, we exploit the fact that such combinatorial pairs are uniform in $\Delta \phi$, and subtract the contribution of pairs from a control region where combinatorial background dominates over the signal pairs.  The region is chosen to be $\abs{\Delta \phi} < \pi/3$, which is symmetric to the back-to-back region with respect to the reaction plane, and thus receives the same contribution from elliptic flow.  Higher order anisotropies are assumed to be negligible for this range in \pt.  Since combinatorial jets are unlikely to pass the b tagging selection, the near-angle contribution is smaller for \cPqb{} dijets than inclusive dijets.

In addition to subtracting the combinatorial component, one also needs to correct for the contribution of signal pairs that are lost when there is a combinatorial jet of higher \pt than the signal partner jet.  To achieve this, an efficiency correction is derived, which is the inverse of the probability that a partner jet of a given \pt was found, i.e., not obscured by a combinatorial jet of larger \pt.  This efficiency is again estimated from data using the small-angle control region, $\abs{\Delta \phi} < \pi/3$.  For a given centrality class, we obtain the spectrum of the highest transverse momentum ($\pt^{\text{max}}$) partner jet in this region in each event.   Assuming that all partner jets in this region are combinatorial, one can derive the probability that a signal partner jet is obscured, as a function of \pt.  This efficiency for detecting the signal partner jet is the cumulative distribution function of this $\pt^\text{max}$ spectrum:

\begin{equation}
 \epsilon(\pt) \equiv 1 - \frac{1}{N} \int_{\pt}^{\infty} \frac{\rd N}{\rd\pt^{\text{max}}} \rd\pt^{\text{max}}.
\label{eq:cdf}
\end{equation}

\begin{figure}[h!]
  \centering
    \includegraphics[width=0.6\textwidth]{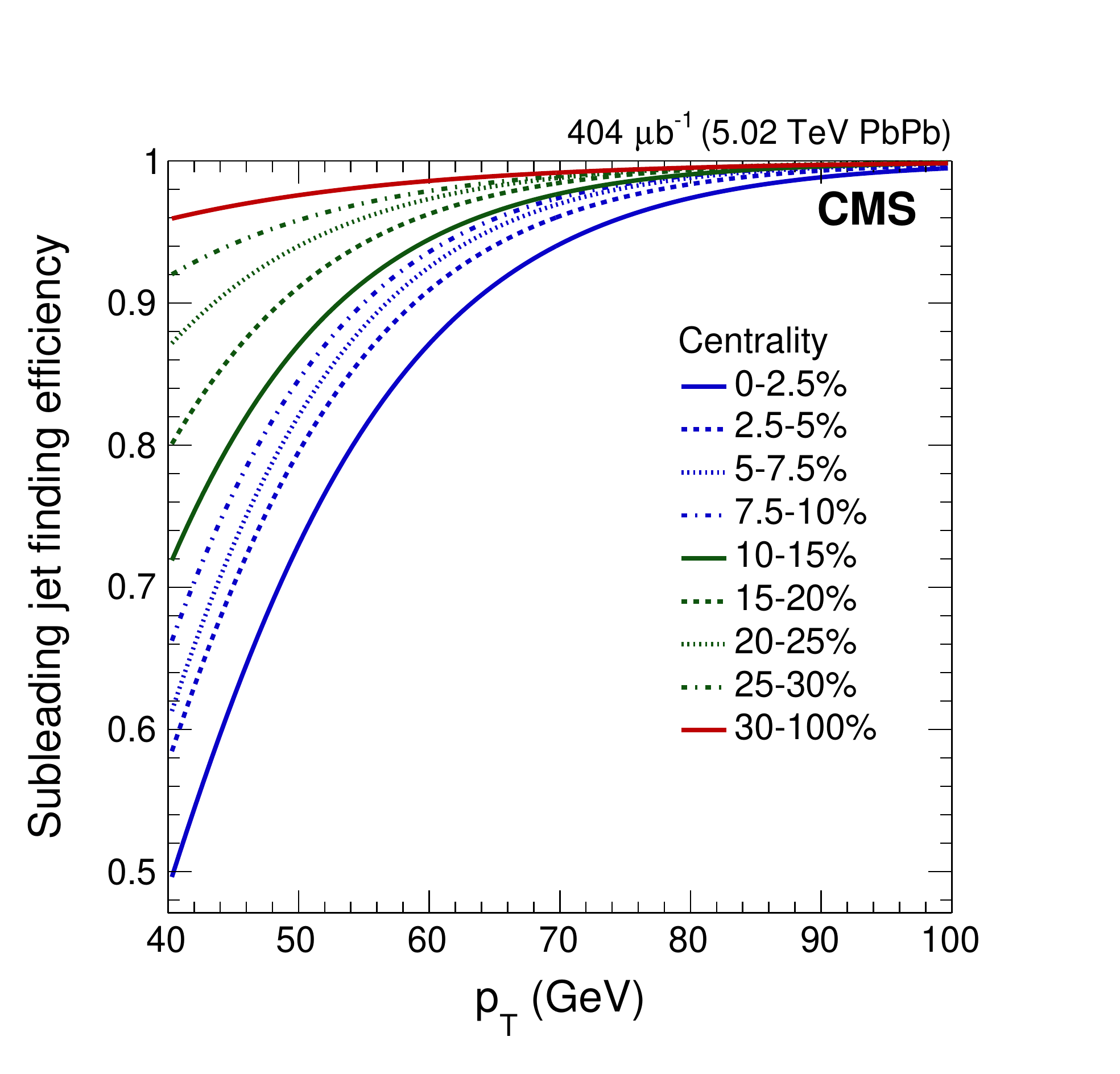}
 \caption[Jet eclipsing correction]{The efficiency of finding a signal partner jet as function of its \pt in \PbPb collisions, as evaluated from the small-angle jet pair control region.  The corrections are shown in the fine centrality bins used in the analysis.
   \label{fig:eclipsecorrection}
 }
\end{figure}

  The efficiency is obtained from a fit to the data in fine bins of centrality, using the Gompertz function, $f(\pt) = \exp[b \exp(c \pt)]$, where $b$ and $c$ are free parameters. The fits obtained are shown in Fig~\ref{fig:eclipsecorrection}.  For each event, the values of $b$ and $c$ for the given centrality are obtained by linear interpolation.  The function with these interpolated parameters is then evaluated at the \pt of the subleading jet.

Although the self-normalized quantities presented in this analysis do not depend on the absolute b tagging efficiency, the relative efficiency as a function of the \pt and $\eta$ must be taken into account.  Corrections are derived from simulation for both the leading and subleading jet.  We also correct for the variation of the b tagging efficiency within the centrality selections presented in this analysis.

In order to probe for quenching or other nuclear effects on the balance distributions, a baseline is constructed using \pp data as a reference.  Since the deterioration of the jet \pt resolution with increasing collision centrality introduces an additional imbalance in the \xj distributions, a direct comparison of \PbPb and \pp measurements does not solely reflect the nuclear modifications. This issue is addressed by smearing the transverse momentum of the jets in \pp data by the amount that corresponds to the additional underlying event fluctuations estimated from \HYDJET simulations that have been tuned to match the underlying event density in \PbPb data.

As in Ref.~\cite{Khachatryan:2016kdb}, the jet \pt resolution is parametrized according the following form, typical for calorimeter energy resolutions.

\begin{equation}
\sigma(\pt)/\pt  = \sqrt{C^2 + S^2/\pt + N^2/\pt^2}
\label{eq:resolution}
\end{equation}

In \pp collisions, the constant ($C$) and stochastic ($S$) terms are 0.06 and 0.8$\sqrt{\GeVns}$, respectively.  In \PbPb collisions the $S$ term has a slightly larger value of 1.0$\sqrt{\GeVns}$, due to the underlying event subtraction.  The noise parameter ($N$) depends on collision centrality, according to $N = 14.82 -\text{centrality (\%)} /5.40 (\GeVns)$.  This term is neglected in \pp collisions.

\section{Systematic uncertainties}
\label{sec:systematics}

The sources of systematic uncertainties in \meanxj\ for the inclusive dijet and \cPqb{} dijet measurements are summarized in Table~\ref{sys:disys} and discussed directly below.

\begin{table}[h!]
\topcaption{ Absolute systematic uncertainties on \meanxj\ for inclusive
(upper sub-table) and b (lower sub-table) dijets.}
\centering
\begin{tabular}{  l  c  c  c  c  c  }
\hline
Source       & \pp      & 30--100\% & 10--30\% & 0--10\% \\
\hline
Combinatorial subtraction    &  \NA       & 0.001    & 0.006    & 0.014 \\
Subleading jet finding &  \NA       & 0.002    & 0.004    & 0.004 \\
Energy scale       & 0.001    & 0.006    & 0.010    & 0.013 \\
Jet resolution         & 0.007    & 0.008    & 0.010    & 0.012 \\ \hline
Total              & 0.007    & 0.010    & 0.016    & 0.023 \\  \hline
Combinatorial subtraction         &  \NA      & 0.008     & 0.008   & 0.008 \\
Subleading jet finding      &  \NA      & 0.002     & 0.004   & 0.004 \\
Tagging efficiency     & 0.002   & 0.003     & 0.003   & 0.009 \\
Signal mistagging           & 0.002  & 0.004     & 0.006   & 0.006 \\
Jet energy scale            & 0.001   & 0.006     & 0.010   & 0.013 \\
Jet resolution         & 0.007    & 0.008    & 0.010    & 0.012 \\ \hline
Total                   & 0.008   & 0.014     & 0.018   & 0.023 \\
\hline
\end{tabular}
\label{sys:disys}
\end{table}

\subsection*{Combinatorial jet pair subtraction}
  The systematic uncertainty in the combinatorial background subtraction in \PbPb collisions is evaluated by varying the contribution of the near-angle control region.  For inclusive dijets, where the near-angle region is dominated by combinatorial jets, the size of the contribution is varied by 30\%, which is sufficient to cover the nonclosure of the subtraction procedure in simulation (the difference between the output of the analysis procedure and the generated input for the simulation).  For \cPqb{} dijets, the number of jet pairs in the near-angle control region is reduced by the b tagging requirement, and is much less centrality-dependent than for inclusive dijets.
Simulations based on \HYDJET embedding show that the dominant contribution in this region corresponds to signal jets from gluon splitting.  We therefore use the entire yield in the near-angle region in \pp data to estimate the systematic uncertainty in the subtraction procedure in \PbPb data.

\subsection*{Subleading jet finding efficiency}
The uncertainty on the efficiency correction for finding the subleading jet is attributed to several effects: a contribution of signal jets in the near-angle control region ($\abs{\Delta \phi} < \pi/3$), the finite centrality binning used and the imperfect description of the Gompertz fit function employed.  The systematic uncertainty associated with these corrections is evaluated from the nonclosure in \HYDJET-embedded simulated samples.

\subsection*{Jet energy scale}
 The uncertainty on the (inclusive) jet energy scale in \pp collisions is evaluated from in-situ studies to be  1\% for the $\eta$ range used in this analysis~\cite{Chatrchyan:2011ds, Khachatryan:2016kdb}.  The same jet energy scale and uncertainty are found to apply to \cPqb{} jets, based on studies of $\cPZ \to \bbbar$.  In-situ studies were also carried out in peripheral \PbPb collisions in Ref.~\cite{Khachatryan:2015lha}, albeit with limited statistics.  A 4\% uncertainty is assigned to cover the observed difference between data and simulation.  The modification of jet fragmentation pattern due to quenching is also a source of systematic uncertainty on the jet energy scale that can be as large as 5\% for the most central collisions~\cite{Chatrchyan:2012gw,Chatrchyan:2014ava}.   Finally, underlying event subtraction leads to an uncertainty in the jet energy scale of up to 2\% for central collisions~\cite{Khachatryan:2015lha,Khachatryan:2016erx}.

To propagate the uncertainties to the \xj distributions, the correlation between the leading and subleading jet energy scales must be taken into account.  For a given jet pair, the ratio \xj is insensitive to an overall shift of the jet energy scale by a multiplicative factor.  Such a shift does, however, effectively change the leading and subleading jet thresholds.  The total correlated shift from the above mentioned sources was estimated to be as large as 6.5\% in central events.  For \cPqb{} dijets, there is an additional systematic uncertainty due to the bias of the b tagging on the jet energy scale, which was evaluated in simulation and found to be 1\% in \pp collisions and 2\% in \PbPb collisions.

There is also a component of the systematic uncertainty that is uncorrelated between the leading and subleading jet.   The subleading jet is also more sensitive to the underlying event subtraction systematics than the leading jet is.   To be conservative, we applied the entire uncertainty of 2\% to the subleading jet, independently of the leading jet.   In addition, to cover the \pt dependence of the modification of the fragmentation pattern due to quenching, the jet energy scale is shifted by a fixed amount, up to 2\GeV in central events.

\subsection*{Jet energy resolution} 
  The uncertainly from the jet resolution is propagated by varying the resolution parametrization in Eq.~\ref{eq:resolution}.  The effect on the \xj distribution is evaluated by applying these alternate smearing parametrizations to particle-level jets.   In \pp collisions, the $C$ and $S$ parameters are varied by 0.02 and 0.2$\sqrt{\GeVns}$, respectively. For \PbPb collisions, in addition, the $N$ term is varied by 2{\GeV}, which covers the difference in underlying event between data and simulation, and the variation of the resolution within the wide centrality bins. Although the results are not unfolded for the resolution effects, the uncertainty is fully included in the data points in order to correctly evaluate any theoretical models that fold in the resolution effects for comparison.

\subsection*{Tagging efficiency (b jets only)}
The tagging efficiency has a fairly flat \pt dependence, such that it has only a mild effect on the observed mean \xj values (\meanxj).  The values of the corrections are varied by 50\% as a conservative estimate of the systematic uncertainty in these corrections.  This is sufficient to cover possible differences in data and simulation observed with studies of the \cPqb{} jet tagging efficiency in control samples in data~\cite{Chatrchyan:2012jua, Chatrchyan:2013exa}.

\subsection*{Mistagging (b jets only)}
The effect of mistagging signal (i.e., not combinatorial) dijets where one or both jets is not associated with a \cPqb{} quark is evaluated by inverting the b tagging selection for both the leading and subleading jets, both independently and simultaneously.  The systematic uncertainty associated with mistagging is based on the imbalance of the inverted selections, taking into account the purity of the \cPqb{} dijet selection in simulation, which is around 85--90\%, depending slightly on centrality.

\section{Results}
\label{sec:results}

\begin{figure}[ht]
  \centering
    \includegraphics[width=0.9\textwidth]{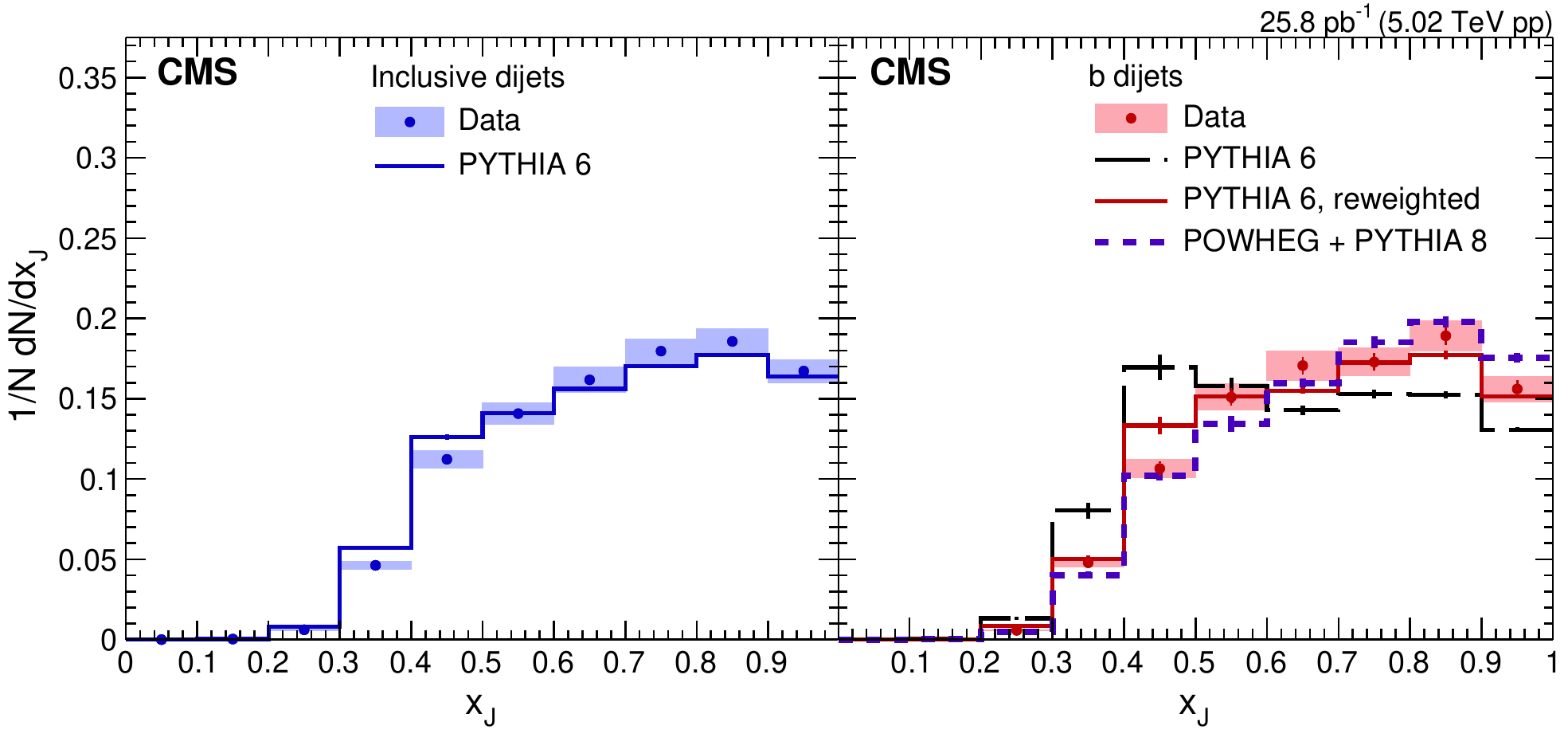}
 \caption[\xj distributions in \pp collisions.]{Distributions of \xj in \pp collisions for inclusive dijets (left) and \cPqb{} dijets (right).  Systematic uncertainties are shown as shaded boxes, while statistical uncertainties are shown as vertical lines. The data are compared to simulations performed using \POWHEG and \PYTHIA, as described in the text. }
\label{fig:xJpp}
\end{figure}

The \pt balance, as quantified by the distribution of \xj, is presented for both inclusive and \cPqb{} dijets.   Both sets of dijets use leading and subleading jet \pt thresholds of 100 and 40\GeV, respectively, selected from jets in $\abs{\eta} < 1.5$.
Figure~\ref{fig:xJpp} shows the distribution in \pp collisions.  The data are compared with simulations performed with \PYTHIA 6, which was found to give an adequate description of the dijet balance for inclusive jets.   The agreement of \PYTHIA 6 with data is notably worse for \cPqb{} dijets, where the simulated distribution is broadened towards imbalanced jet pairs.   This broad feature is not observed in the \cPqb{} dijet data, which instead shows an \xj distribution that resembles that of inclusive dijets.
It was found that improved agreement could be obtained by reweighting the contributions of heavy-flavor production processes in \PYTHIA 6, a procedure which is discussed in the Appendix.  The reweighted distribution is also shown in Fig~\ref{fig:xJpp}.   Finally, the data are also compared to simulations based on next-to-leading order matrix elements, as encoded in the hvq package~\cite{Frixione:2007nw} of the \POWHEG \textsc{box}~\cite{Alioli:2010xd} (v2) generator.  Hadronization in the \POWHEG method~\cite{Frixione:2007vw,Nason:2004rx} is performed by matching the matrix elements to parton showers, which in this case are generated with \PYTHIA 8.212~\cite{Sjostrand:2007gs},  tune CUETP8M1~\cite{Khachatryan:2015pea}.   The \POWHEG + \PYTHIA 8 simulations are found to give a good description of the \cPqb{} dijet data.

\begin{figure}[ht]
  \centering
    \includegraphics[width=0.9\textwidth]{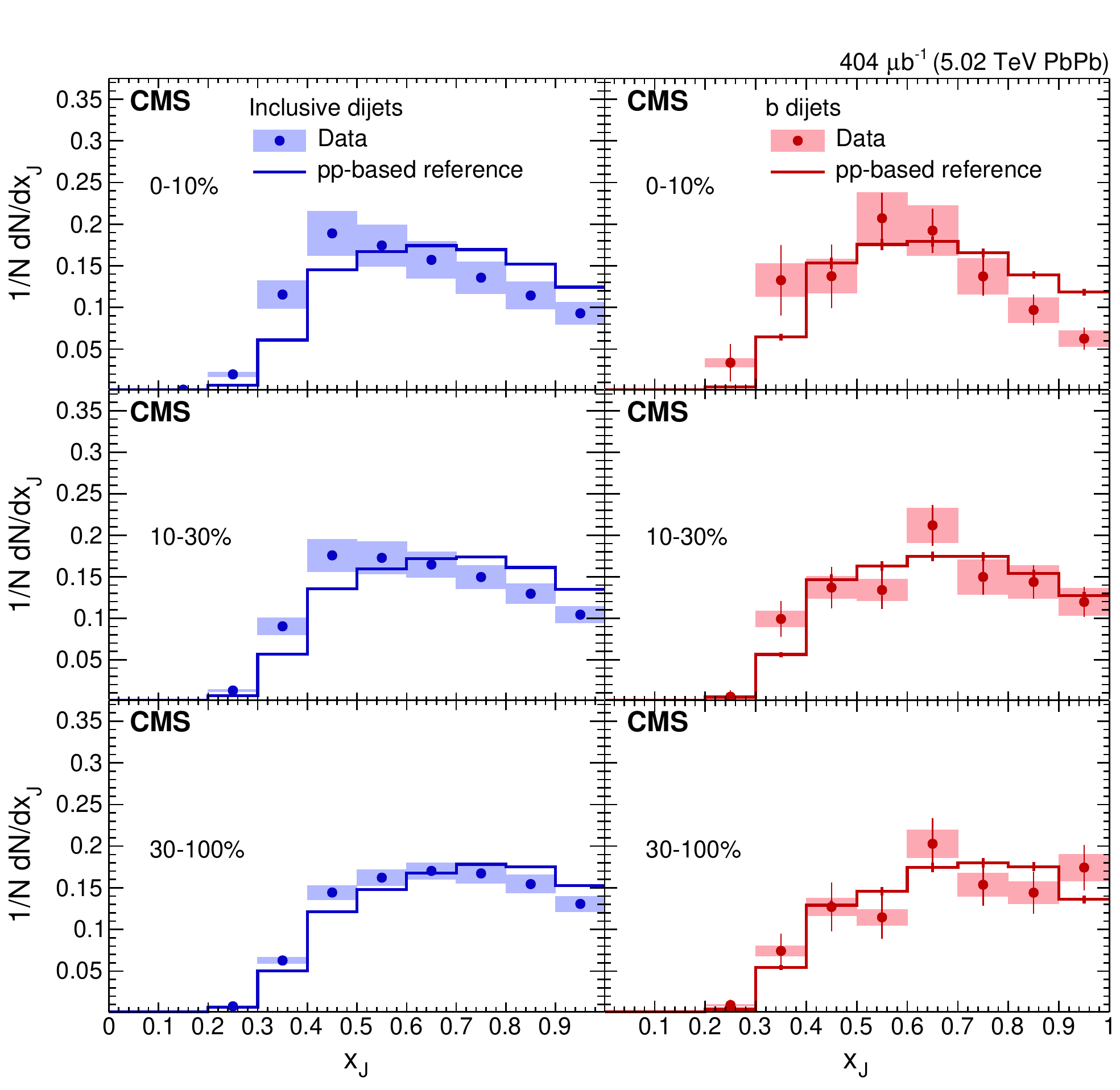}
 \caption[\xj distributions in \PbPb collisions.]{Distributions of \xj in \PbPb collisions for inclusive dijets (left) and \cPqb{} dijets (right).  Systematic uncertainties are shown as shaded boxes, while statistical uncertainties are shown as vertical lines.  The top, middle and bottom rows show the 0--10, 10--30 and 30--100\% centrality selections, respectively. The data are compared to a reference obtained by smearing \pp according to the jet resolution for the given centrality class, as described in the text. }
\label{fig:xJPbPb}
\end{figure}

\begin{figure}[ht]
  \centering
    \includegraphics[width=1.0\textwidth]{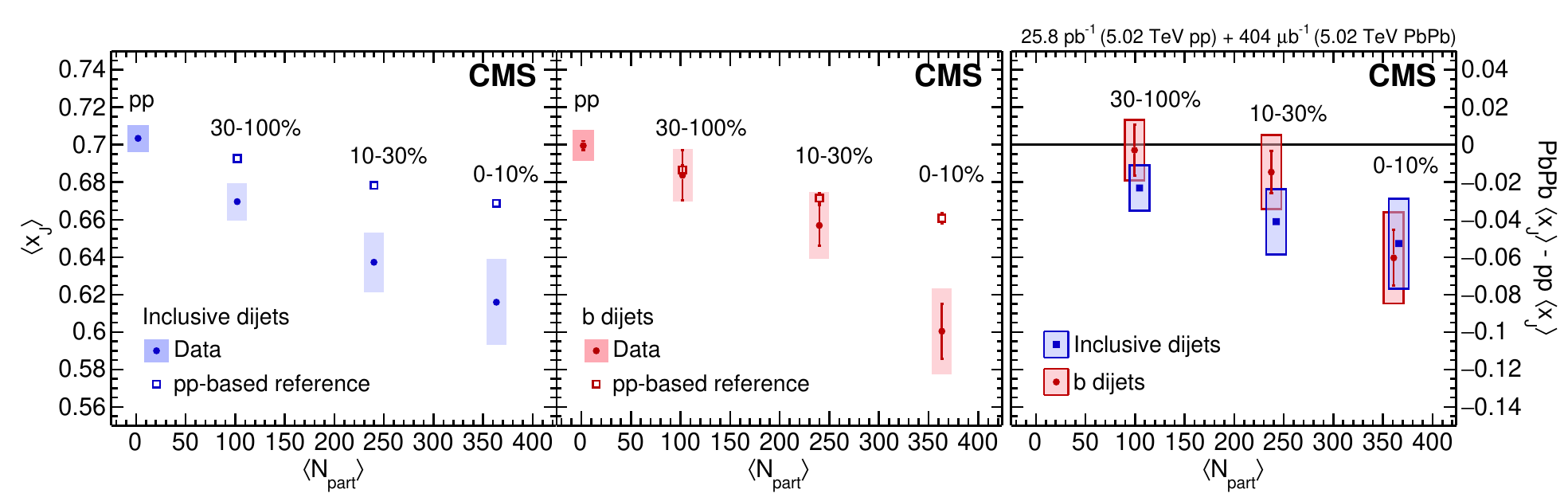}
 \caption{\meanxj\ for inclusive (left) dijets and \cPqb{} dijets (center) in \pp collisions and for different centrality selections of \PbPb collisions. The right panel shows the difference in the \meanxj\ values between \PbPb and the smeared \pp reference. Systematic uncertainties are shown as shaded boxes, while statistical uncertainties are shown as vertical lines.  }
\label{fig:xjmean}
\end{figure}

 Figure~\ref{fig:xJPbPb} shows the \xj distributions for inclusive dijets and \cPqb{} dijets for three different centrality selections of \PbPb collisions.  Here the data are compared to the reference obtained from \pp data by smearing the \pt of each jet according to a parametrization of the resolution for the given centrality class.
 Figure~\ref{fig:xjmean} shows the \meanxj\ values from these distributions, as well as the difference between the \meanxj\ in \PbPb and the smeared \pp reference.  The data are plotted as a function of the number of participants estimated from a Monte Carlo Glauber model~\cite{Miller:2007ri,Alver:2008zza}.  The number of participants is weighted by the number of collisions to account for the hard scattering bias within each bin.  Both the inclusive dijet and \cPqb{} dijet data show a tendency towards increasing imbalance with increasing centrality.  While the reference data also become more imbalanced because of resolution effects, the magnitude of the effect is clearly smaller.   The effect is understood to result from jet quenching, as observed in previous inclusive dijet results~\cite{Chatrchyan:2011sx,Chatrchyan:2012nia}. For inclusive dijets, a clear quenching signal is observed already for the 30--100\% centrality bin. For \cPqb{} dijets, on the other hand, the imbalance is compatible with the \pp reference in the 30--100\% bin.  In the 10--30\% bin, the \cPqb{} dijet data point lies between the inclusive dijet one and the \pp reference, within two standard deviations of both.  Only in the most central bin (0--10\%) is the \cPqb{} dijet quenching significant at the level of about three standard deviations, with a value close to that observed for inclusive dijets.

\section{Conclusions}
\label{sec:conclusions}

In this paper, transverse momentum (\pt) correlations of \cPqb{} quark jet pairs (\cPqb{} dijets) have been measured in \PbPb collisions for the first time, and compared to results from \pp collisions.  In \pp collisions, a similar \pt balance distribution was observed for inclusive dijets and \cPqb{} dijets.  For the latter case, \POWHEG was found to give a better description than \PYTHIA 6 alone (without reweighting), suggesting that next-to-leading order effects are important for the modeling of this observable.  This should be taken into consideration for models of parton energy loss in nucleus-nucleus collisions, which often use leading order calculations or generators as input.  In \PbPb collisions the net \pt imbalance was observed to be larger in the most central collisions for \cPqb{} dijets, as had already been observed for inclusive dijets.  This effect can be understood to originate from the energy loss of partons in the quark-gluon plasma.  In the most central bin, the observed quenching effect is of comparable magnitude for \cPqb{} dijets and for inclusive dijets, the latter of which contains a mixture of quark and gluon jets.  Insofar as parton energy loss is thought to depend on the type of parton that initiates the parton shower, this measurement can place constraints on the underlying dynamics of the interaction of the parton with the quark-gluon plasma.

\begin{acknowledgments}
We congratulate our colleagues in the CERN accelerator departments for the excellent performance of the LHC and thank the technical and administrative staffs at CERN and at other CMS institutes for their contributions to the success of the CMS effort. In addition, we gratefully acknowledge the computing centers and personnel of the Worldwide LHC Computing Grid for delivering so effectively the computing infrastructure essential to our analyses. Finally, we acknowledge the enduring support for the construction and operation of the LHC and the CMS detector provided by the following funding agencies: BMWFW and FWF (Austria); FNRS and FWO (Belgium); CNPq, CAPES, FAPERJ, and FAPESP (Brazil); MES (Bulgaria); CERN; CAS, MoST, and NSFC (China); COLCIENCIAS (Colombia); MSES and CSF (Croatia); RPF (Cyprus); SENESCYT (Ecuador); MoER, ERC IUT, and ERDF (Estonia); Academy of Finland, MEC, and HIP (Finland); CEA and CNRS/IN2P3 (France); BMBF, DFG, and HGF (Germany); GSRT (Greece); OTKA and NIH (Hungary); DAE and DST (India); IPM (Iran); SFI (Ireland); INFN (Italy); MSIP and NRF (Republic of Korea); LAS (Lithuania); MOE and UM (Malaysia); BUAP, CINVESTAV, CONACYT, LNS, SEP, and UASLP-FAI (Mexico); MBIE (New Zealand); PAEC (Pakistan); MSHE and NSC (Poland); FCT (Portugal); JINR (Dubna); MON, RosAtom, RAS, RFBR and RAEP (Russia); MESTD (Serbia); SEIDI, CPAN, PCTI and FEDER (Spain); Swiss Funding Agencies (Switzerland); MST (Taipei); ThEPCenter, IPST, STAR, and NSTDA (Thailand); TUBITAK and TAEK (Turkey); NASU and SFFR (Ukraine); STFC (United Kingdom); DOE and NSF (USA).

\hyphenation{Rachada-pisek} Individuals have received support from the Marie-Curie program and the European Research Council and Horizon 2020 Grant, contract No. 675440 (European Union); the Leventis Foundation; the A. P. Sloan Foundation; the Alexander von Humboldt Foundation; the Belgian Federal Science Policy Office; the Fonds pour la Formation \`a la Recherche dans l'Industrie et dans l'Agriculture (FRIA-Belgium); the Agentschap voor Innovatie door Wetenschap en Technologie (IWT-Belgium); the F.R.S.-FNRS and FWO (Belgium) under the "Excellence of Science - EOS" - be.h project n. 30820817; the Ministry of Education, Youth and Sports (MEYS) of the Czech Republic; the Council of Science and Industrial Research, India; the HOMING PLUS program of the Foundation for Polish Science, cofinanced from European Union, Regional Development Fund, the Mobility Plus program of the Ministry of Science and Higher Education, the National Science Center (Poland), contracts Harmonia 2014/14/M/ST2/00428, Opus 2014/13/B/ST2/02543, 2014/15/B/ST2/03998, and 2015/19/B/ST2/02861, Sonata-bis 2012/07/E/ST2/01406; the National Priorities Research Program by Qatar National Research Fund; the Programa Severo Ochoa del Principado de Asturias; the Thalis and Aristeia programs cofinanced by EU-ESF and the Greek NSRF; the Rachadapisek Sompot Fund for Postdoctoral Fellowship, Chulalongkorn University and the Chulalongkorn Academic into Its 2nd Century Project Advancement Project (Thailand); the Welch Foundation, contract C-1845; and the Weston Havens Foundation (USA).
\end{acknowledgments}

\bibliography{auto_generated}

\clearpage
\appendix
\section{Appendix}
\label{sec:appendix}

\begin{figure}[h!]
  \centering
    \includegraphics[width=0.49\textwidth]{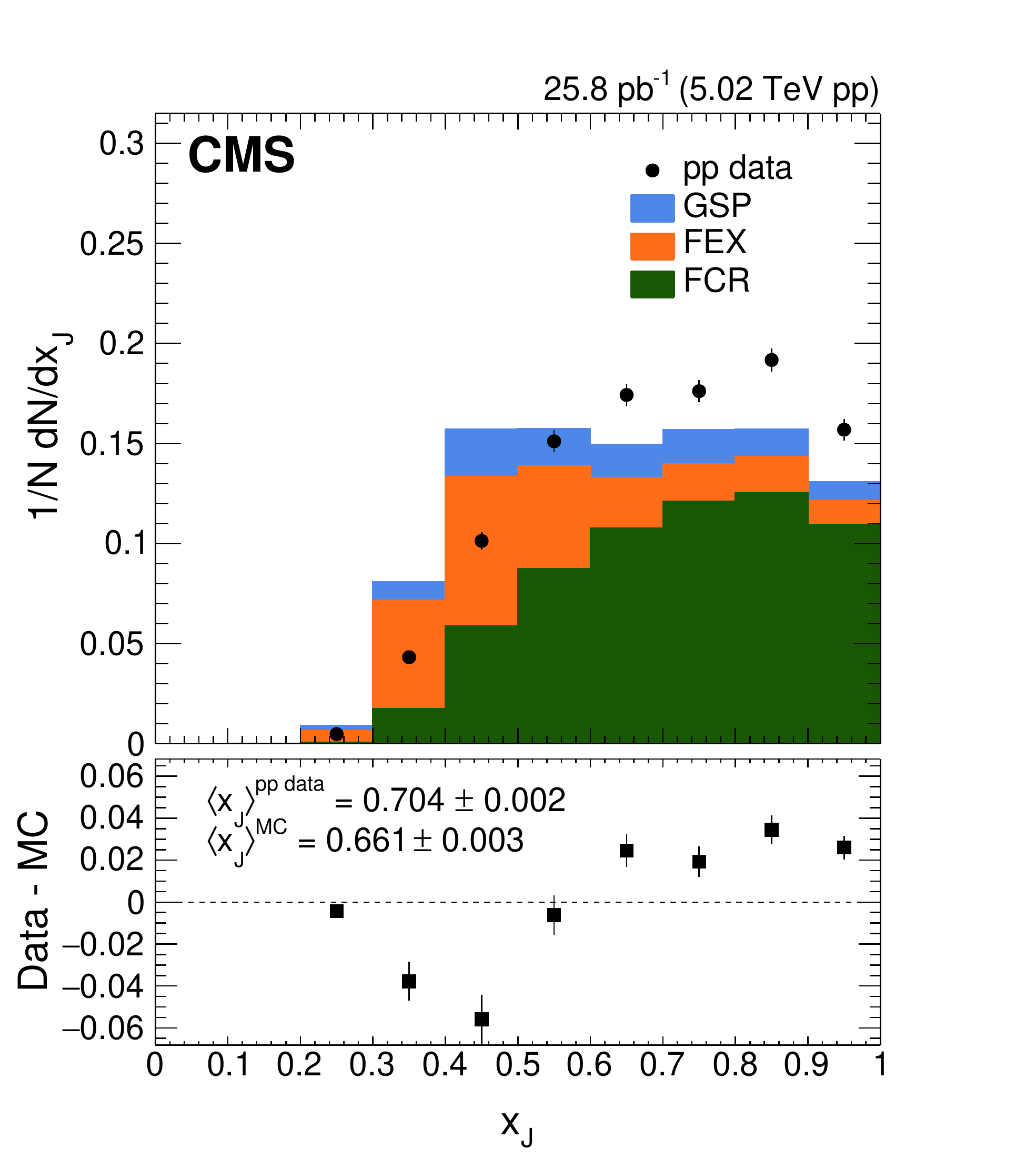}
    \includegraphics[width=0.49\textwidth]{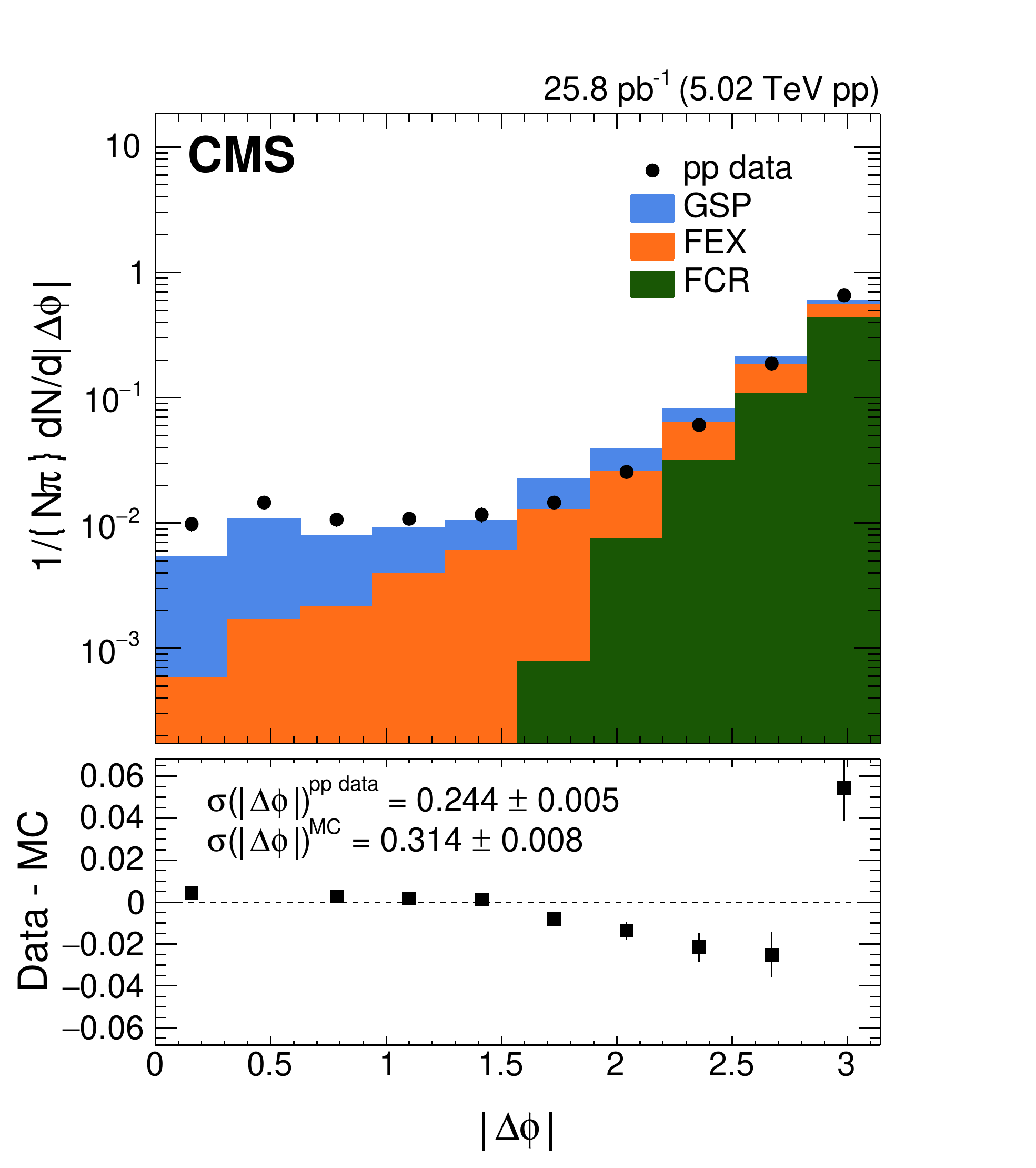}
 \caption[\xj and \dphi  distributions in the leading-subleading analysis before flavor process reweighting.]{
The distributions of \xj (left) and \dphi  (right) in \pp collisions before flavor process reweighting. Data are shown in solid points, while the stacked histograms show the contributions of different processes in \PYTHIA 6 (see text for details).  The bottom set of panels show the difference between data and simulation (MC).
}
\label{fig:LS_xjdphi_noWeight}
\end{figure}

Whereas tunes of \PYTHIA 6 used to compare to LHC data give a reasonable description of the dijet balance for inclusive jets (e.g., in Ref.~\cite{Chatrchyan:2012nia}), they fail to adequately describe the angular and  \pt correlations between \cPqb{} jet pairs for the kinematic range probed by this measurement, as shown in the \xj and \dphi  distributions in Fig.~\ref{fig:LS_xjdphi_noWeight}.  To understand the nature of this discrepancy simulated \bbbar{} events are separated into three categories, depending on the number of outgoing b (or \cPaqb) quarks in the $2 \to 2$ hard scattering. In the flavor creation process (denoted FCR), both of the outgoing particles are \cPqb{} quarks.  The gluon fusion reaction ($\cPg\cPg \to \bbbar$) dominates, with a small contribution from quark-antiquark annihilation ($\cPq\cPaq \to \bbbar$).  In the flavor excitation process (FEX) only one of the outgoing particles is \cPqb{} quark.  In this case, a virtual gluon in one of the protons has split into a $\bbbar$ pair and one of the \cPqb{} quarks enters the hard scattering.  In the process referred to here as gluon splitting (GSP), neither of the outgoing particles is a \cPqb{} quark.  The parent may be a gluon that participates in the hard scattering or a gluon that appears elsewhere in the event, for example in a parton shower.

The discrepancy of \PYTHIA 6 with the data is driven by the poor modeling of the FEX contribution, which tends to give \cPqb{} dijet pairs that are too asymmetric in \pt.  This discrepancy was already noted by the CDF Collaboration~\cite{Acosta:2004nj}, and may be understood as follows. The partner \cPqb{} quark in the FEX process is treated as initial-state radiation.  The \PYTHIA 6 tunes require large initial-state radiation to describe \TeV scale collider data.  However, such tunes over-predict the probability that the partner \cPqb{} quark at mid-rapidity and enters the kinematic selections used in this analysis.   While an improved modeling of this process can be achieved by softening the initial-state radiation, this would have an impact on other observables, in particular the overall dijet \pt balance.  Instead, the contribution of the three heavy-flavor production modes are reweighted according to the following procedure.   Three exclusive categories of events are defined, using jets within $\abs{\eta} < 1.5$:

\begin{itemize}
\item The two highest \pt jets are \cPqb-tagged and back-to-back ($\abs{\dphi_{1,2}} > 2\pi/3$);
\item The first and third highest \pt jet are \cPqb-tagged and back-to-back ($\abs{\dphi_{1,3}} > 2\pi/3$);
\item The first and third highest \pt jet are \cPqb-tagged and nearby ($\abs{\dphi_{1,3}} < \pi/3$).
\end{itemize}

\begin{figure}[h!]
  \centering
    \includegraphics[width=0.49\textwidth]{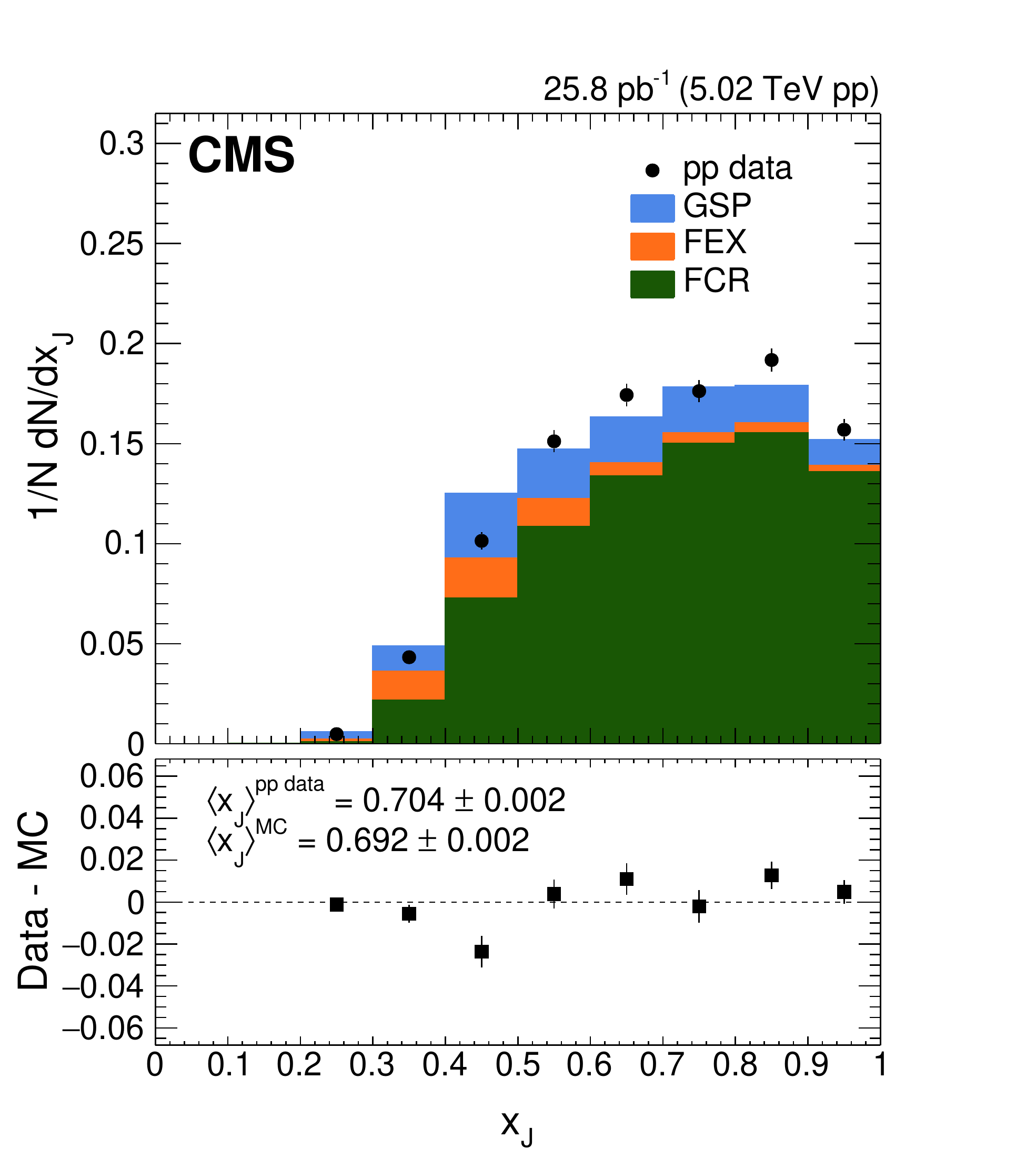}
    \includegraphics[width=0.49\textwidth]{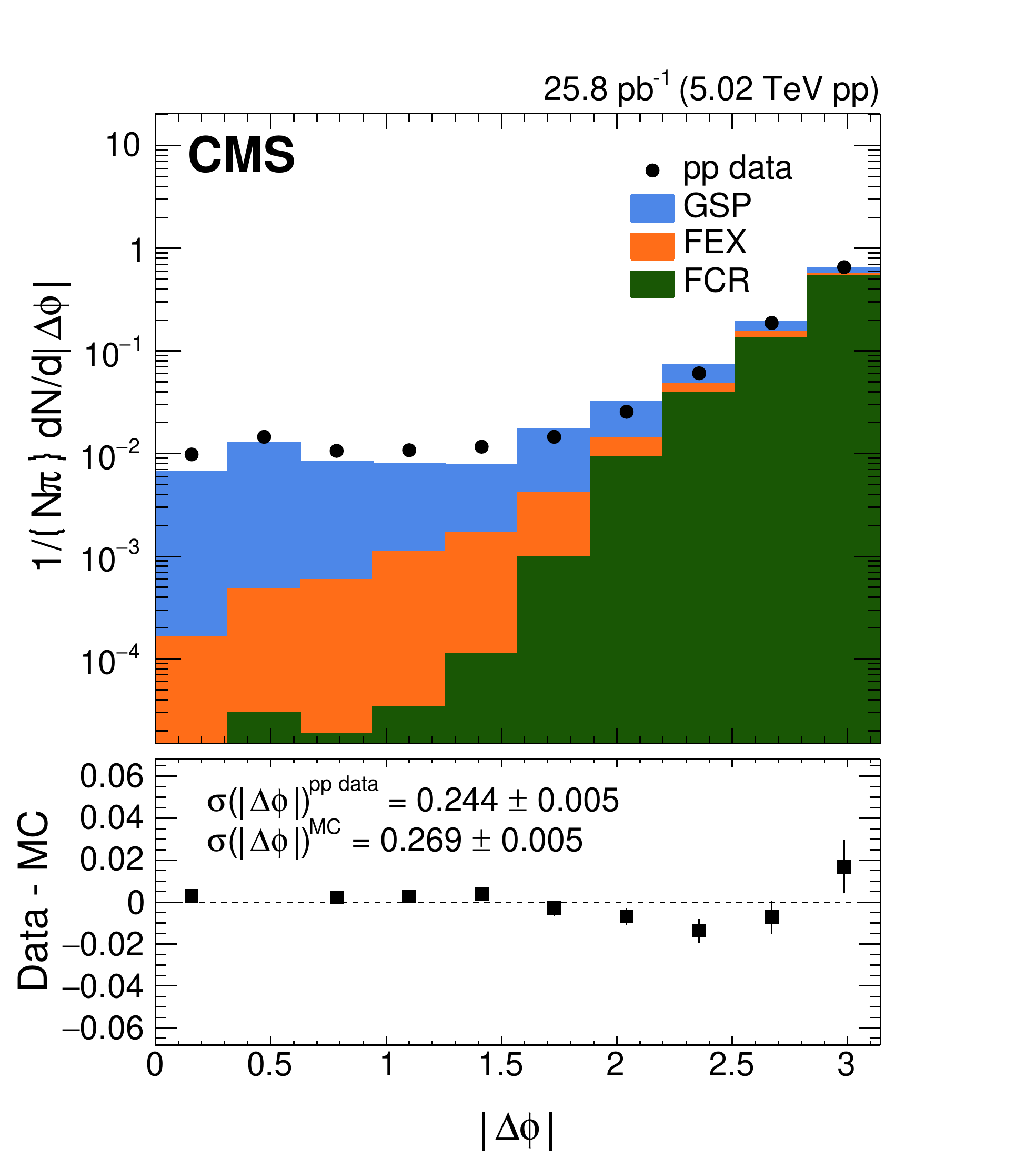}
 \caption[\xj and \dphi  distributions after flavor process reweighting]{
The distributions of \xj (left) and \dphi  (right) in \pp collisions after flavor process reweighting. Data are shown in solid points, while the stacked histograms show the contributions of different processes in \PYTHIA 6 (see text for details).   The bottom set of panels show the difference between data and simulation (MC).
}
\label{fig:LS_xjdphi_Weight}
\end{figure}

In simulation, these categories are found to be dominated by FCR, FEX, and GSP events, respectively.  The contribution of each process in simulation is reweighted such that the relative abundance of these three categories of events are the same as in data. The relative contributions of the three heavy-flavor production sub-processes to these categories are shown in Table.~\ref{tab:procjet}.  Also shown in Table.~\ref{tab:procjet} are the relative occurrences of the three categories in data and simulation.  Finally, Table~\ref{tab:reweightresult} shows the relative contribution of the three production processes to selected \cPqb{} dijets before and after the reweighting.   The contribution of the FCR process to the selected \cPqb{} dijet events is found to be at the level of 70\% in \PYTHIA 6  after the reweighting procedure is applied. Figure~\ref{fig:LS_xjdphi_Weight} shows the improved agreement of the \xj and \dphi  distributions between data and simulation after reweighting.

\begin{table}[h]
\centering
\topcaption{Relative contributions of the three heavy-flavor production sub-processes in \PYTHIA 6 to the jet pair categories, as well as the relative abundance of the three categories in data and simulation. }
\begin{tabular}{  l  c  c  c  c c }
\hline
Category & FCR & GSP & FEX & Data & Simulation\\
\hline
$\abs{\dphi_{1,2}} > 2\pi/3$ & 57\% & 17\% & 26\% & 56\% & 46\%\\
$\abs{\dphi_{1,3}}> 2\pi/3$ & 11\% & 27\% & 62\% & 37\% & 49\%\\
$\abs{\dphi_{1,3}} < \pi/3$ & 0\% & 83\% & 17\% &   7\% &  5\%\\
\hline
\end{tabular}
\label{tab:procjet}
\end{table}

\begin{table}[h]
\centering
\topcaption{Contributions of the three production processes to selected dijets in \PYTHIA 6 before and after reweighting.}
\begin{tabular}{  c c  c  }
\hline
Process & Default & Reweighted \\
\hline
FCR & 53\% & 70\% \\
FEX & 33\% & 9\% \\
GSP & 14\% & 21\% \\
\hline
\end{tabular}
\label{tab:reweightresult}
\end{table}

\cleardoublepage \section{The CMS Collaboration \label{app:collab}}\begin{sloppypar}\hyphenpenalty=5000\widowpenalty=500\clubpenalty=5000\textbf{Yerevan Physics Institute,  Yerevan,  Armenia}\\*[0pt]
A.M.~Sirunyan, A.~Tumasyan
\vskip\cmsinstskip
\textbf{Institut f\"{u}r Hochenergiephysik,  Wien,  Austria}\\*[0pt]
W.~Adam, F.~Ambrogi, E.~Asilar, T.~Bergauer, J.~Brandstetter, E.~Brondolin, M.~Dragicevic, J.~Er\"{o}, A.~Escalante Del Valle, M.~Flechl, M.~Friedl, R.~Fr\"{u}hwirth\cmsAuthorMark{1}, V.M.~Ghete, J.~Grossmann, J.~Hrubec, M.~Jeitler\cmsAuthorMark{1}, A.~K\"{o}nig, N.~Krammer, I.~Kr\"{a}tschmer, D.~Liko, T.~Madlener, I.~Mikulec, E.~Pree, N.~Rad, H.~Rohringer, J.~Schieck\cmsAuthorMark{1}, R.~Sch\"{o}fbeck, M.~Spanring, D.~Spitzbart, A.~Taurok, W.~Waltenberger, J.~Wittmann, C.-E.~Wulz\cmsAuthorMark{1}, M.~Zarucki
% ...
\vskip\cmsinstskip
% ...
1:~~Also at Vienna University of Technology, Vienna, Austria\\
% ...

\end{sloppypar}
\end{document}